\newcommand{\beq}{\begin{equation}}
\newcommand{\eeq}{\end{equation}}
\newcommand{\barr}{\begin{eqnarray}}
\newcommand{\earr}{\end{eqnarray}}
\documentclass{jabad}

 \usepackage{graphicx}

\begin{document}

% ====================================================================
% Title, authors, abstract, keywords and AMS codes
% ====================================================================

\title{A physics pathway to the Riemann hypothesis}

% If your title is too long to fit on the running head:
% \title[short title]{Title of your article}

% If your title is too long to fit on one line:
% \title[short title]{title first line  \\ title second line}

% Nombre abreviado de autores para poner en las cabeceras de pagina
 \newcommand{\shortauthor}{{G. Sierra}}
%afiliacion completa de los autrores
 \author{Germ\'an Sierra$^1$\footnote{german.sierra@uam.es}}

\begin{center}
 {$^1$ Instituto de F\'{\i}sica Te\'{o}rica UAM/CSIC, Facultad de Ciencias,\\
  Universidad Aut\'onoma de Madrid, 28049 Cantoblanco, Spain}
\end{center}
%\author{ Rafael Rodriguez Trias}

%\footnote{${(1)}$ julio@unizar.es,  ${(2)}$ mrios@unizar.es}

% Two authors with the same address

% \begin{address}
%   first author and second author \\
%   first line of the address \\
%   second line \\
%   last line \\
%   \texttt{first author's email} and \texttt{second author's email}
% \end{address}

% Two authors with different addresses: first option

% \begin{address}
%   first author \\
%   first line of the address \\
%   second line \\
%   last line \\
%   \texttt{first author's email}
% \end{address}
%
% \begin{address}
%   second author \\
%   first line of the address \\
%   second line \\
%   last line \\
%   \texttt{second author's email}
% \end{address}

%\address{Departamento de F\'{\i}sica Te\'{o}rica, Facultad de Ciencias,}
%\centerline{Departamento de F\'{\i}sica Te\'{o}rica, Facultad de Ciencias,}
%\centerline{Universidad de Zaragoza, 50009 Zaragoza, Spain}

% If there are two authors:
% \author{First author and second author}

% If there are three or more authors:
% \author{First author, second author,...,nth author and last author}

% If the authors' names do not fit on one line:
% \author[abbreviated list]{first line \\ second line}
% In the optional argument put something like first author et al.

\begin{abstract}
 We present a brief review of the spectral approach to the Riemann hypothesis, 
 according to which the imaginary part of the non trivial zeros of the zeta function are the 
 the eigenvalues of the Hamiltonian of a quantum mechanical system. 
 \end{abstract}

\KeysAndCodes{Riemann hypothesis, zeta function,quantum mechanics, quantum chaos}{02.10.De, 05.45.Mt, 11.10.Hi}

% ====================================================================
% Section 1
% ====================================================================

\section{Introduction}

The Riemann hypothesis (RH) is the statement that the complex zeros
 of the classical zeta function all 
have imaginary  part equal to $1/2$. It was first suggested by Riemann in his famous memoir
in 1859 \cite{Ri}. 
 The RH is important for its connection with the distribution of prime numbers 
 \cite{Bo,Co}. The average number of primes less than a given number $x$, which is denoted as 
$\pi(x)$, behaves asymptotically as $x/\log x$. This statement is called the Prime Number Theorem
(PNT) and was proved independently by Hadamard and de La Vall\' ee-Poussin 
in 1896 \cite{Ed,Ti}. The truth of the RH implies 
that the fluctuations  of $\pi(x)$, around its average value, behaves asymptotically as $x^{1/2} \log x$, which also gives the  best possible bound for the error of the PNT.   The RH is not an isolated property of a particular function, but it holds for the Dirichlet L-functions, for curves over finite 
fields, etc. It is expected that a proof of the RH for the zeta function, will be generalizable to other 
L-functions as well. However the consensus is that some  key idea is required
for this goal. 

One of the most promising pathways for a proof of the RH was suggested by Polya and Hilbert
around 1910, but never published apparently. The suggestion is that there exist a selfadjoint 
operator $H$, whose spectrum contains the imaginary part of the Riemann zeros. The selfadjointness
of such an operator would  inmediately prove the RH: 

\beq
{\rm If} \; \;  \zeta\left( \frac{1}{2} + i E_n \right) = 0,  \; \; \forall n 
\Longrightarrow H \; |\psi_{n}  \rangle = E_n \; | \psi_{n}  \rangle \Longrightarrow  E_n = E_n^*,
\; \; \forall n 
\label{1}
\eeq

In the latter equation one must exclude the trivial Riemann zeros, so a most
appropiate formulation fo the problem is to use the $\xi(s)$ function, defined as \cite{Ed}

\beq
\xi(s) = \frac{1}{2} s ( s-1) \pi^{- s/2} \Gamma \left(  \frac{s}{2} \right) \zeta(s)
\label{1b}
\eeq

\noindent 
whose zeros
are those of $\zeta(s)$ except the trivial ones. Polya and Hilbert conjecture is known as the
spectral approach to the RH, and,  as we shall see later on, it is supported by several
"phenomenological" results and heuristic arguments, which suggest that the operator $H$ is the quantum Hamiltonian of a physical  system. This is one of the reasons why the RH has attracted  the interest of physicist working in disciplines apparently unrelated to Number Theory (see \cite{Wa} for an extensive
list of references concerning several approaches to the RH).

 Assuming the RH (i.e. $E_n^* = E_n, 
\forall n$),  one can define a 
diagonal operator $H$ whose entries are $E_n$, 
but nothing is learned from this construction. 
Eq.(\ref{1}), implies that $H$ must encode in itself, the zeta function $\zeta(1/2 + i E)$,
without assuming the truth of the RH, which will be a consequence of its selfadjointness. If 
$\zeta(1/2 + i E)$, or $\xi(1/2 + i E)$, where a polynomial in $E$, this encoding could be realized by a finite dimensional
matrix $H$ whose characteristic polynomial were proportional to $\zeta(1/2 + i E)$. 
The Euler product formula of the zeta function, in terms of the prime numbers, implies
 that $H$ must also know about these numbers. Hence,  the relation found by Riemann, between
prime numbers and zeros of the zeta function, will be justified from the common 
dynamical origin of these quantities. The precise mathematical formulation of these
relations is given by the,  so called, trace formulas in Number Theory and  Quantum Chaos \cite{Ca}.  
The simplicity of the definition of the zeta function, as the series $\zeta(s) = \sum_{n=1}^\infty 1/n^s
\; \; ({\rm Re} \; s > 1)$, has lead some researches, as Berry, to suggest that the Hamiltonian
$H$ may have a subtle but simple definition, allowing the 
observation of the Riemann zeros as spectral lines in an experimental set up \cite{Be}. 
The existence of such a "Riemann calculator" would place 
Number Theory in the realm of Quantum Mechanics with far reaching consequences. 
We shall briefly summarize below some phenomenologial and heuristic hints that 
support the spectral approach to the RH. 

\section{ Selberg's trace formula (1956)}

Consider a compact Riemann surface with negative curvature. This surface
can be constructed as the complex upper plane divided by a discrete subgroup of the modular
group $PSL(2, {\bf R} )$,  and it is equipped  with the Poincar\' e metric.
 A classical problem is to determine the lengths 
$\ell_p$  of the primitive  periodic orbits (p.p.o.), that is, the geodesics on this surface. A quantum problem 
is to find the spectrum of the Laplace-Beltrami operator $\Delta= y^2( \partial_x^2 + \partial_y^2)$, 

\beq
- \Delta \; \psi_n(x,y) = E_n \; \psi_n(x,y), \qquad E_n = \frac{1}{4} + k_n^2. 
\label{2}
\eeq

Selberg's trace formula establishes a relation between the momenta $k_n$
and the length of the geodesics $\ell_p$ \cite{Se}

\beq
\sum_n h(k_n) = \frac{ \mu(D)}{4 \pi} \int_{- \infty}^\infty
dk \; k \,  h(k) \,  {\rm tanh} (\pi k) + \sum_{\rm p.p.o.} \ell_p \sum_{n=1}^\infty
\frac{ g(n \ell_p)}{ 2 \sinh( n \ell_p/2)}
\label{3}
\eeq

\noindent
where $h(k)$ is a test function, $g(k)$ its the Fourier transform and $\mu(D)$ is the area
of the fundamental domain $D$ describing the Riemann surface.  
Selberg also defined a zeta function
in terms of the lenghts $\ell_p$ as

\beq
Z(s) = \prod_{\rm p.p.o.}  \prod_{m=0}^\infty ( 1 - e^{- \ell_p (s + m)})
\label{4}
\eeq

\noindent
in close analogy to  the Euler' s product formula of the zeta function \cite{Ed}:

\beq
\zeta(s) = \prod_{p} \frac{1}{ 1 - p^{-s}}, \qquad {\rm Re} \; s > 1 
\label{5}
\eeq

\noindent 
where the product is over all the prime numbers $p$. Selberg zeta function
satisfies a RH which can be proved. The trivial zeros of $Z(s)$ are 
$s_n = - n \;(n=0,1, \dots)$,  and the non trivial ones are $s_n = \frac{1}{2} + i k_n$. 
Since $k_n$ are real numbers, any complex zero of $Z(s)$ lies on the 
line $ {\rm Re} \; s = 1/2 $.  The functions $\zeta(s)$ and $Z(s)$ both 
satisfy functional equations that relate
their  values to $\zeta(1-s)$ and  $Z(1-s)$, respectively. Finally,  Selberg' s trace formula
is reminiscent to the Riemann-Weil explicit  formula relating  the prime numbers $p$
and the imaginary part of the Riemann zeros $\gamma_n$ \cite{We}:

\beq
\sum_n h(\gamma_n) =  \int_{- \infty}^\infty
\frac{dk}{2 \pi}   h(k) \,  \frac{\Gamma'}{\Gamma} \left( 
\frac{ 1}{4} \frac{i k}{2} \right)+ h(\frac{i}{2}) + h(-\frac{i}{2}) - \log \pi \; g(0)
  - 2  \sum_{\rm p} \log p \sum_{n=1}^\infty
p^{- n/2}  g(n \log p)
\label{6}
\eeq

\noindent 
where the notations are as in eq.(\ref{3}). A comparison between eqs.(\ref{3}) and
(\ref{6}) suggests that prime numbers and primitive geodesics are in one-to-one correspondence, such that: $\ell_p \leftrightarrow \log p$. This correspondence also underlines
the Quantum Chaos approach to the RH reviewed later on. However, 
the analogy between these two formulas fails in two respects: i) the term 
$1/(2 \sinh(n \ell_p/2))$, with the identification $\ell_p = \log p$, 
only converges  to $p^{- n/2}$ for large values of  $p$, and ii) the factor of -2
in the last term of eq.(\ref{6}) as compared to the factor one, in the last term of eq.(\ref{3}). The difference in signs of the two terms finds an explanation in Connes spectral
realization of Riemann zeros (see below). Finally, we observe that the imaginary part
of the Riemann zeros, $\gamma_n$, seem to correspond to the momenta $k_n$ 
rather than to the eigenenergies $E_n$. This suggest that the Riemann Hamiltonian
is probably related to a first order linear operator, as we shall see in the discussion of the
$H = xp$ Hamiltonian.

\section{Random matrix theory and Quantum Chaos (70's-80' s)}

In 1973  Montgomery, asuming the RH, proved that the imaginary part of the Riemann zeros,
$\gamma_n$, where distributed at random, according the gaussian unitary ensemble distribution 
(GUE) of Random Matrix Theory (RMT) \cite{Mo}.  This result found strong numerical confirmation by Odlyzko in the 80' s, who computed  trillions of Riemann zeros ( near the $10^{12}$ -th zero 
and near the $10^{20}$- th zero) \cite{Od}.  These phenomenological 
findings  means that the statistical properties of the
Riemann zeros are similar to those of the eigenvalues of large hermitean matrices, in particular
the property of level repulsion. There are three universality classes of random matrices
corresponding  to orthogonal (GOE), hermitean (GUE) and symplectic matrices (GSE). The 
GUE statistics corresponds to random systems where time reversal is broken, which
gives a strong indication that the Riemann Hamiltonian H must break this symmetry. 

A further step along this direction was taken by Berry,  who noticed a formal analogy
between the fluctuations of the Riemann zeros 
and the fluctuations of the energy levels of quantum chaotic system around their
average values \cite{Be}.  The latter fluctuations are given by the semiclassical Gutwiller formula,

\beq
N_{QC, \rm fl}(E) = \frac{1}{\pi} \sum_\gamma \sum_{m=1}^\infty 
\frac{ \sin( m E T_\gamma)}{ 2 m \sinh( m \lambda_\gamma/2)}
\label{7}
\eeq

\noindent
where $E$ is an eigenenergy, $\gamma$ is a primitive periodic orbit, 
$T_\gamma$ its period and $\lambda_\gamma$ its Lyapunov exponent. 
The sum over $m$ corresponds to the repetitions of the primitive orbits. 
The fluctuation part of the Riemann zeros is given by 

\beq
N_{R, \rm fl}(E) = - \frac{1}{\pi} \sum_p  \sum_{m=1}^\infty 
\frac{ \sin( m E \log p)}{ m p^{m/2} }
\label{8}
\eeq

\noindent 
where the sum is over the prime numbers $p$. 
Comparing  (\ref{7}) and  (\ref{8}),   Berry conjectured the 
existence of a classical chaotic Hamiltonian whose primitive periodic orbits, $\gamma$, 
 would be labelled  by the prime numbers $p$,  with periods  $T_p = \log p$,  and instability exponents 
$\lambda_p =  \pm \log p$.  Moreover, since each orbit is counted once, the Hamiltonian must 
break time reversal (otherwise there would be a factor $2/\pi$,  in front of eq. (\ref{8}) instead 
of $1/\pi$).  This analogy is reminiscent to the one existing between Selberg  trace formula
and the Riemann-Weil formula, and also suffers from a "sign problem" and "asymptotic problem"
as observed before.  The connection with Quantum Chaos also explained some numerical discrepancies found by Odlyzko between RMT and the statistics of zeros for long range spectral correlations, 
which are due to the shortest periodic orbits, where universality no longer holds. 
They were explained by Berry , Keating and Bogomolny \cite{Be3,BoK}. All these results put on a more
firm basis the Polya-Hilbert conjecture giving further clues on the structure of the dynamical
system behind the Riemann zeros.

\section{The Hamiltonian $H = x p$  (1999)}

In 1999 Berry and Keating, and Connes suggested that the Riemann zeros are related to the
classical Hamiltonian
$H_{\rm cl} = x p$, where $x$ and $p$ are the position and momenta of a particle
movin in 1D \cite{BK,Con}.   The classical trayectories of this Hamiltonian are hyperbolas
in the phase space

\beq
x(t) = x_0 \; e^t, \qquad p(t) = p_0 \; e^{-t}
\label{9}
\eeq

\noindent
and therefore unbounded, which would then imply a continuous spectrum rather
than a discrete spectrum associated to the Riemann zeros. The connection with the latter
arises in two possible ways depending on two different regularizations of the
phase space. Berry and Keating introduced a minimal length $\ell_x$ and a minimal
momenta $\ell_p$, whose product is the Planck quantum $\ell_x \ell_p = 2 \pi \hbar$ \cite{BK}.
In terms of these quantities,  they imposed $|x| \geq \ell_x$ and $|p| \geq \ell_p$, so that
the trayectories are now bounded. The semiclassical number of states is given by the area
below  the hyperbola  and above the boundaries $|x| = \ell_x$ and $|p| =  \ell_p$, and it 
is given, in units $\hbar = 1$, by

\beq
n_{\rm BK} (E) = \frac{A_{\rm BK}}{2 \pi}  = 
 \frac{E}{2 \pi}  \left(  \log  \frac{E}{2 \pi}  - 1 \right) + 1
\label{10}
\eeq

\noindent
Rather surprisingly, this result coincides, asymptotically,  with the average number of Riemann
zeros up to a height $E$ in the critical strip (i.e. $ 0 < \Re s < 1, 0 < \Im s < E)$. The constant
 in Riemann' s   formula is actually 7/8,  which can be obtained by taking into account 
a Maslow phase contributing  $-1/8$  to eq. (\ref{10}), due to the fact that the particle only travels
 one quadrant of the phase space. Unfortunately, eq.(\ref{10}) has remained so far heuristic
 since it is not supported by a quantum mechanical model (see later). 
 
 Connes regularization is based on the restrictions $|x| \leq  \Lambda$ and $|p| \leq \Lambda$,
 where  $\Lambda$ is a cutoff, which is taken to infinity at the end of the calculation \cite{Con}. The semiclassical
 number of states is computed as before yielding,

\beq
n_{\rm Co} (E) = \frac{A_{\rm Co}}{2 \pi}  =  \frac{E}{2 \pi } \log
  \frac{ \Lambda^2}{ 2 \pi} - 
 \frac{E}{2 \pi}  \left(  \log  \frac{E}{2 \pi}  - 1 \right) 
\label{11}
\eeq

\noindent 
The first term on the RHS of this formula diverges in the limit  $\Lambda \rightarrow \infty$,
which corresponds to a continuum of states. The second term is minus the
average number of Riemann zeros, which according to Connes, become 
missing spectral lines in the continuum. This is the, so called, "absortion" spectral interpretation
of the Riemann zeros, as opposed to the standard "emission" spectral interpretation where
they form a discrete spectrum. The minus sign in eq.(\ref{11}) could also be related to the minus
sign in the trace formulas discussed earlier. Connes interpretation has however two drawbacks.
First of all,  the average number of Riemann zeros is not fully obtained in eq(\ref{11}). The
term $E/2 \pi \log 1/2 \pi$, actually cancells between the first and second summands in this
formula. Other objection  is that the  second term in (\ref{11}) is simply a finite size correction of 
discrete  energy  levels, where no lines are missing, and the same remains true in the
continuum limit. 

The $H=xp$ model was modified in references \cite{Si}, 
adding a non local interaction suggested by a relation of this model
to a BCS model of superconductivity with a cyclic renormalization group. 
 The spectrum of this interacting $xp$-model
  is a continuum where the Riemann zeros are embedded as bound states.
This result reconciles the Berry-Keating and Connes spectral interpretations.
However,  the non locality of the interaction implies that the Hamiltonian
has no classical limit, and consequently its relation to classical chaotic dynamical 
systems remains unclear. Moreover the prime numbers do not appear in this
construction, which as we saw in previous sections,  is an important
ingredient of the trace formulas.

A different route to $H=xp$ was suggested in reference \cite{ST}
which will bring us to more familiar territores in Physics.

\section*{Landau levels and Riemann zeros (2008)}

Let us consider a charged particle moving in a plane under the action of a perpendicular
magnetic field and an electrostatic potential with a saddle shape \cite{ST}. The Langrangian
describing the dynamics is given, in the Landau gauge, by

\beq
{\cal L}  =  \frac{\mu}{2} ( \dot{x}^2 + \dot{y}^2 ) - \frac{e B }{c} \dot{y} x - e \lambda x y 
\label{12}
\eeq

\noindent
where $\mu$ is the mass, $e$ the charge, $B$ the magnetic field, $c$ the speed
of light and $\lambda$ a coupling constant  parameterizing  the electrostatic potential. There are
two normal modes with  real,  $\omega_c$, and imaginary, $\omega_h$, angular
frequencies, describing  cyclotronic and a hyperbolic motions respectively. In the limit
where $\omega_c >> \omega_h$, only the Lowest Landau Level (LLL)  is relevant and the effective Lagrangian becomes

\beq
{\cal L}_{\rm eff}   =  p \dot{x} - |\omega_h |  x p, \qquad p = \frac{\hbar y}{\ell^2}, \qquad
\ell = \left(  \frac{ \hbar c}{ e B} \right)^{1/2},  \qquad |\omega_h| \sim \frac{ \lambda c}{B} 
\label{13}
\eeq

\noindent  
where $ \ell$  is the magnetic length, which is proportional to 
the radius of the cyclotronic orbits in the LLL.  The coordinates $x$ and $y$, which
commute in the 2D model, after the proyection to the LLL become canonical
conjugate variables,  and the effective Hamiltonian coincides with  the $xp$ Hamiltonian
introduced by Berry, Keating and Connes, where the energy is measured in units of
$\hbar |\omega_h|$. This  realization of the $xp$ Hamiltonian allow us to interpret the
semiclassical quantization of these authors in the language of the Landau model. 
In particular,  the semiclassical counting of states in the $xp$ model follows
from the counting of quantum fluxes in a certain area of the $x-y$ plane. 
If the plane is infinite, then the number of states in the LLL  will is also infinite. To have a finite
number of states we put the particle into  a box: $ |x| < L, |y| < L$, which reproduces  Connes regularizations conditions.  The number of semiclassical states with an energy
between $0$ and $E$ is given by

\beq
n_{\rm sm} (E)   =  \frac{E}{2 \pi } \log
  \frac{ L^2}{ 2 \pi \ell^2} - 
 \frac{E}{2 \pi}  \left(  \log  \frac{E}{2 \pi}  - 1 \right) 
\label{14}
\eeq

\noindent 
which agrees with Connes eq. (\ref{11}). The classical energy is given by $E = x y/\ell^2$
(in units of $\hbar |\omega_h|$), and it attains its maximum value at $E_{\rm max} = L^2/\ell^2$.
Plugging this value into (\ref{14}) yields $n_{\rm sm}(E_{\rm max}) = L^2/2 \pi \ell^2$, which 
is the number of quantum fluxes in the first quadrant. This semiclassical results
can be derived from the quantization of the model. Indeed,  the energy 
levels follows from the identification of the wave function at the boundaries
$x=L$ and $y = L$ (up to a phase)

\beq
\psi_E(x, L) = e^{ i x L/\ell^2} \psi_E(L, x) \Longrightarrow
\frac{ \Gamma \left( \frac{1}{4} + \frac{i E}{2} \right)}{ 
\Gamma \left( \frac{1}{4} - \frac{i E}{2} \right)}  \left(  \frac{ L^2}{ 2 \ell^2} \right)^{ - i E} = 1
\label{15}
\eeq

\noindent 
Taking the logarithm on the RHS of (\ref{15}),  one gets the smooth part of the  Riemann 
formula,  whose asymptotic expansion  coincides with eq.  (\ref{14}). The 2D formulation
of the $xp$ model allows one to convert the hyperbolic orbits into periodic orbits
by means of the boundary condition  (\ref{15}). It would be extremely interesting
to derive Berry-Keating regularization in the Landau version of the model.  To achieve  this
goal,  one must inject the particle approaching the boundary at $y = \ell_y$, back
to the boudary at $x= \ell_x$, so that the orbits become periodic. 
The Berry-Keating semiclassical arguments suggest that the spectrum will be 
discrete and associated to the smooth Riemann zeros. Preliminary results
suggest that this possibility can indeed be realized. Of course the main problem that 
remains is the construction of the Hamiltonian giving rise to the exact Riemann zeros.
It is not clear at the moment wether this can be done, but preliminary results
suggest that the higher Landau levels may play a role \cite{ST}. In any case, the Landau
model formulation of the $H=xp$ Hamiltonian provides a promissing new avenue
where to explore the fascinating problem of a physical interpretation of the Riemann zeros,
and perhaps a physicist proof of the RH.

% ====================================================================
% Acknowledgements
% ====================================================================

\ack 
To the memory of Julio Abad, who was a good man in every sense of the word.
I wish to thanks the members of the Departamento de F\'{\i}sica Te\'orica
of the University of Zaragoza, and specially Manuel Asorey,  for the opportunity to contribute to the hommage of Julio.
This work has been supported by the Spanish CICYT grant  FIS2004-04885
and the  ESF Science Programme  INSTANS 2005-2010. 

% ====================================================================
% References
% ====================================================================


\begin{thebibliography}{99}
%\bibliographystyle{acmurl}
%\bibliography{bibmodel}


\bibitem{Ri} B. Riemann, "Ueber die Anzahl der Primzahlen unter einer gegebenen Gr\"osse," Monat. der K\"onigl.  Preuss. Akad. der Wissen. zu Berlin aus der Jahre 1859, 671 (1860). 

\bibitem{Bo} E. Bombieri, "Problems of the millennium: The Riemann hypothesis," Clay Mathematics 
Institute (2000),  http://www.claymath.org/millennium/Riemann Hypothesis/Official Problem Description.pdf. 

\bibitem{Co} B. Conrey, "The Riemann hypothesis," AMS Notices, 341 (2003). 

\bibitem{Ed}H. M. Edwards, Riemann's zeta function, New York: Academic Press (1974). 

\bibitem{Ti} E. C. Titchmarsh, The theory of the Riemann zeta function, Oxford: Clarendon Press (1986). 


\bibitem{Wa}M. R. Watkins, http://www.maths.ex.ac.uk/ ?mwatkins/zeta/physics.htm. 


\bibitem{Ca} "Frontiers in Number Theory, Physics, 
and Geometry I On Random Matrices, Zeta Functions, 
and Dynamical Systems", Eds: P. Cartier, B. Julia, P. 
Moussa, P. Vanhove. Springer Verlag, Berlin, 2006.

\bibitem{Be} Berry M V 1986 Riemann's zeta function: a model for quantum chaos? Quantum Chaos and Statistical Nuclear 
Physics ed T H Seligman and H Nishioka vol 263, 1-17 


\bibitem{Se}A. Selberg, "Harmonic analysis and discontinuous groups in weakly symmetric Riemannian spaces 
with applications to Dirichlet series," J. Indian Math. Soc. 20, 47 (1956). 
See also D. Hejhal, "The Selberg Trace Formula for PSL(2, R)", Vol. I, Springer Lecture Notes 548 
(1976). 

\bibitem{We}A. Weil, "Sur les formules explicites de la th\'eorie des nombres premiers,"
Meddelanden Fran Lunds Univ. Mat. Sem., 252 (1952); 


\bibitem{Mo} H. L. Montgomery, "The pair correlation of zeros of the zeta function," in Analytic Number 
Theory, vol. 24 of AMS Proceedings of Symposia in Pure Mathematics, 181 (1973). 

\bibitem{Od} A. M. Odlyzko, "On the distribution of spacings between zeros of the zeta function," Math. 
Computation 48, 273 (1987). 


\bibitem{Be3} M.V.  Berry,  "Semiclassical formula for the number variance of the Riemann zeros",  Nonlinearity 1 399-407 (1988). 


\bibitem{BoK} E. B. Bogomolny and J. P. Keating "Gutzwiller' s trace formula and spectral statistics: beyond the diagonal  approximation",  Phys. Rev. Lett. 77 1472-5 (1996). 


\bibitem{BK} M. V. Berry and J. Keating, "H = xp and the Riemann zeros"  in Supersymmetry and trace 
formulae: chaos and disorder, I. V. Lerner and J. P. Keating, eds., New York: Plenum (1999). 
 "The Riemann zeros and eigenvalue asymptotics," SIAM Review 41, 236 (1999) and references 
therein. 

\bibitem{Con} A. Connes, "Trace formula in noncommutative geometry and the zeros of the Riemann zeta 
function," Sel. Math., New Ser. 5, 29 (1999) [arXiv:math/9811068]. 
For more details consult the recent book A. Connes and M. Marcolli, Noncommutative geometry, 
quantum Þelds, and motives, AMS (2008). 


\bibitem{Si} G. Sierra, "H = x p with interaction and the Riemann zeros," Nucl. Phys. B 776, 327 (2007) 
[arXiv:math-ph/0702034];  "The Riemann zeros and the cyclic Renormalization Group," J. Stat. Mech. 0512, P006  (2005) [arXiv:math.nt/0510572]; 
"A quantum mechanical model of the Riemann zeros," New J. Phys. 10, 033016 (2008) 
[arXiv:0712.0705 [math-ph]]; 


\bibitem{ST} G. Sierra and P. K. Townsend, "Landau levels and Riemann zeros," Phys. Rev. Lett. 101, 110201  (2008) [arXiv:0805.4079 [math-ph]]. 

  \end{thebibliography}
\end{document}